\documentclass[aps,prl,reprint,groupedaddress,showpacs]{revtex4-1}

\usepackage{graphicx}

\begin{document}


\title{Electric Field Control of the Verwey Transition and Induced Magnetoelectric Effect in Magnetite}


\author{Jared J. I. Wong}
\affiliation{Department of Physics and Astronomy, University of California, Riverside,
CA 92521, USA}

\author{Adrian G. Swartz}
\affiliation{Department of Physics and Astronomy, University of California, Riverside,
CA 92521, USA}

\author{Renjing Zheng}
\affiliation{Department of Physics and Astronomy, University of California, Riverside,
CA 92521, USA}

\author{Wei Han}
\affiliation{Department of Physics and Astronomy, University of California, Riverside,
CA 92521, USA}

\author{Roland K. Kawakami}
\email[]{roland.kawakami@ucr.edu}
\affiliation{Department of Physics and Astronomy, University of California, Riverside,
CA 92521, USA}


\date{\today}

\begin{abstract}
We incorporate single crystal Fe$_3$O$_4$ thin films into a gated device structure and demonstrate the ability to control the Verwey transition with static electric fields.  
The Verwey transition temperature ($T_V$) increases for both polarities of the electric field, indicating the effect is not driven by changes in carrier concentration. Energetics of induced electric polarization and/or strain within the Fe$_3$O$_4$ film provide a possible explanation for this behavior. Electric field control of the Verwey transition leads directly to a large magnetoelectric effect with coefficient of 585 pT m/V.
\end{abstract}


\maketitle


Electric field control of magnetic and metal-to-insulator transitions in highly correlated 
materials has generated great interest both scientifically and
technologically \cite{ahn:2003,asamitsu:1997,lee:2008,lottermoser:2004,zheng:2004}.
Magnetite (Fe$_3$O$_4$) is a highly correlated material that undergoes the well-known 
Verwey transition with sharp changes in the electric, magnetic, and structural 
properties \cite{verwey:1939,verwey:1941,walz:2002}. Following over 
seven decades of intense research since its discovery 
in 1939, the origin of the Verwey transition as a charge order-disorder transition has 
become clear experimentally just in the last few 
years \cite{ding:2008,garcia:2009,huang:2006,lorenzo:2008,rozenberg:2006,
senn:2011,zhou:2010}. 
In this Letter, we report the
surprising observation of electric field control of the Verwey transition in Fe$_3$O$_4$ thin films.
An electric field applied by electrostatic gates is found to stabilize the Verwey structure 
and increase the transition temperature ($T_V$). Furthermore, electrical control of $T_V$ leads
 to a new mechanism for generating a magnetoelectric effect. We obtain a magnetoelectric 
coefficient of 585 pT m/V, which is one of the largest for a single-phase material. These 
results provide an alternative approach to advanced electronics, information processing, 
and spintronics \cite{lee:2010,li:2010,nagasawa:2007,vaz:2009}.

Magnetite is the oldest known magnetic material (lodestone) and the Verwey 
transition at $T_V\sim$ 120 K is one of the first examples of a metal-to-insulator transition in 
which the insulating phase is generated by electron-electron 
correlations \cite{verwey:1939,verwey:1941,walz:2002}. The Verwey 
transition also includes sharp changes to the 
structure and magnetization \cite{walz:2002}.  Cooling 
through $T_V$, the unit cell structure changes from inverse-spinel to monoclinic and the 
magnetization exhibits a sharp decrease below $T_V $\cite{walz:2002}. This coupling between the 
electronic, magnetic, and structural properties makes magnetite very attractive for 
exploring tuning of magnetoelectric behavior 
in correlated materials which could lead to novel functionality.  

Previous demonstrations of electric field control of magnetic and metal-to-insulator transitions 
have resulted from various effects including current-induced breakdown of the insulating 
state \cite{asamitsu:1997,lee:2008,wang:2010}, field-induced changes of 
carrier concentration \cite{lottermoser:2004,vaz:2010}, field-induced 
strain generated by growth on 
piezoelectric substrate (i.e. composite 
system) \cite{ryu:2001,ryu:2002,zheng:2004,yan:2010}. 
The effect reported here is distinct from 
these previous categories. Unlike current-induced breakdown which is a highly 
non-equilibrium process, the present effect produces a true change in the equilibrium 
phase transition. We also find that this effect is not due to changes in carrier 
concentration, as shown by a symmetric dependence of $T_V$ on gate 
voltage. Finally, this effect does not 
rely on external strain provided by adjacent layers. Thus, the electric field control of 
the Verwey transition represents a new 
type of electric field control in a highly correlated material.

\begin{figure}
\includegraphics[width=80mm]{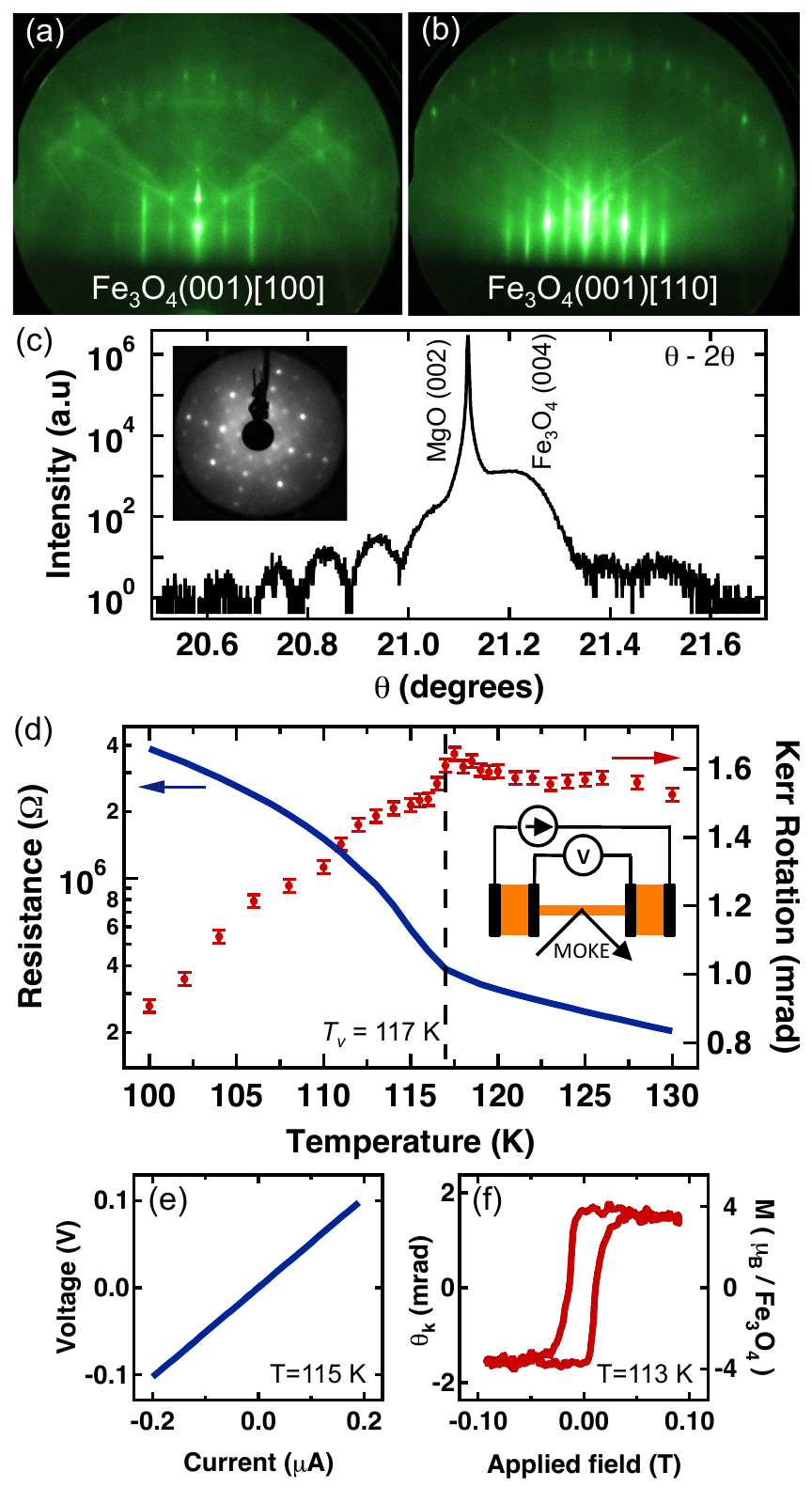}
\caption{\label{fig:materials} Characterization of Fe$_3$O$_4$ thin films. (a) and (b) are RHEED 
patterns for 50 nm Fe$_3$O$_4$ on MgO(001) along the [100] and [110] in-plane directions,
respectively. (c) HRXRD $\theta-2\theta$ scans measured around the location of the MgO(002)
peak with Kiessig fringes.  Inset: LEED pattern with incident energy of 160 eV.  (d) Temperature
dependence of resistance measured by four-point probe (blue) and magnetization measured by
MOKE (red). The vertical dashed line indicates the Verwey transition. Inset: geometry for the
resistance and magnetization measurements. (e) \emph{I}-\emph{V} curve for Fe$_3$O$_4$
channel at 115 K.  (f) MOKE hysteresis loop for Fe$_3$O$_4$ at 113 K. 
The right axis shows absolute
magnetization based on SQUID measurements on corresponding samples.}
\end{figure}

Fe$_3$O$_4$ films of 50 nm thickness are grown on double-side polished MgO(001) substrates 
using reactive molecular beam epitaxy (MBE)  in 
ultrahigh vacuum (UHV) with a base pressure of $1\times10^{-10}$ torr. MgO substrates are first rinsed 
with de-ionized (DI) water. After loading into the MBE chamber, substrates are annealed 
at 600$^\circ$C for 45 minutes.  A 10 nm MgO buffer layer is grown at 350$^\circ$C via electron 
beam (e-beam) deposition from an MgO 
source \cite{wong:2010}.  Next, the Fe$_3$O$_4$ layer is grown at 
200$^\circ$C by depositing elemental Fe in a 
molecular oxygen partial pressure of $1.2\times10^{-7}$ torr. The 
Fe is evaporated from a thermal effusion cell at a rate of $\sim$0.13 nm/min (for pure Fe).  
The single-crystal structure is verified 
through \emph{in situ} reflection high energy electron diffraction (RHEED) and low energy 
electron diffraction (LEED), as shown in 
Figs. \ref{fig:materials}a, \ref{fig:materials}b, and \ref{fig:materials}c inset. $\theta-2\theta$ high 
resolution x-ray diffraction (HRXRD) scans exhibit a Fe$_3$O$_4$(004) peak near the 
MgO(002) substrate peak (Fig. \ref{fig:materials}c). Kiessig interference fringes indicate atomically 
smooth interfaces and verify the film thickness.

Electrical properties of Fe$_3$O$_4$ films are characterized using standard dc four-point 
probe measurements (Figure \ref{fig:materials}d inset). Resistance values are obtained from 
current-voltage (\emph{I}-\emph{V}) curves, which 
exhibit linear dependence (Figure 1e) above 70 K. 
The temperature dependence of resistance 
(Figure \ref{fig:materials}d, blue) exhibits a metal-to-insulator 
transition with a substantially higher resistance below 117 K, indicating the Verwey transition.
Temperature dependence curves are measured as a function of increasing temperature, with the temperature stabilized for 10 min before a measurement is taken. For 
each temperature, measurements are repeated to ensure the temperature is stable.

The magneto-optic Kerr effect (MOKE), with laser beam incident through the transparent 
MgO substrate, is used to characterize the magnetic properties of the Fe$_3$O$_4$ films 
(812 nm wavelength, \emph{p}-polarized, 45$^\circ$ angle of incidence). Figure 1f shows a typical 
longitudinal MOKE hysteresis loop that exhibits large remanence and sharp 
magnetization reversal. The right hand axis of Fig. 1f displays the corresponding 
magnitude of the magnetization based on superconducting quantum interference device 
(SQUID) magnetometry. The temperature dependence of the MOKE signal (Figure 1d, red) 
exhibits a suppression of magnetization for temperatures below $T_V$. This behavior is 
characteristic of the Verwey transition.

\begin{figure}
\includegraphics[width=80mm]{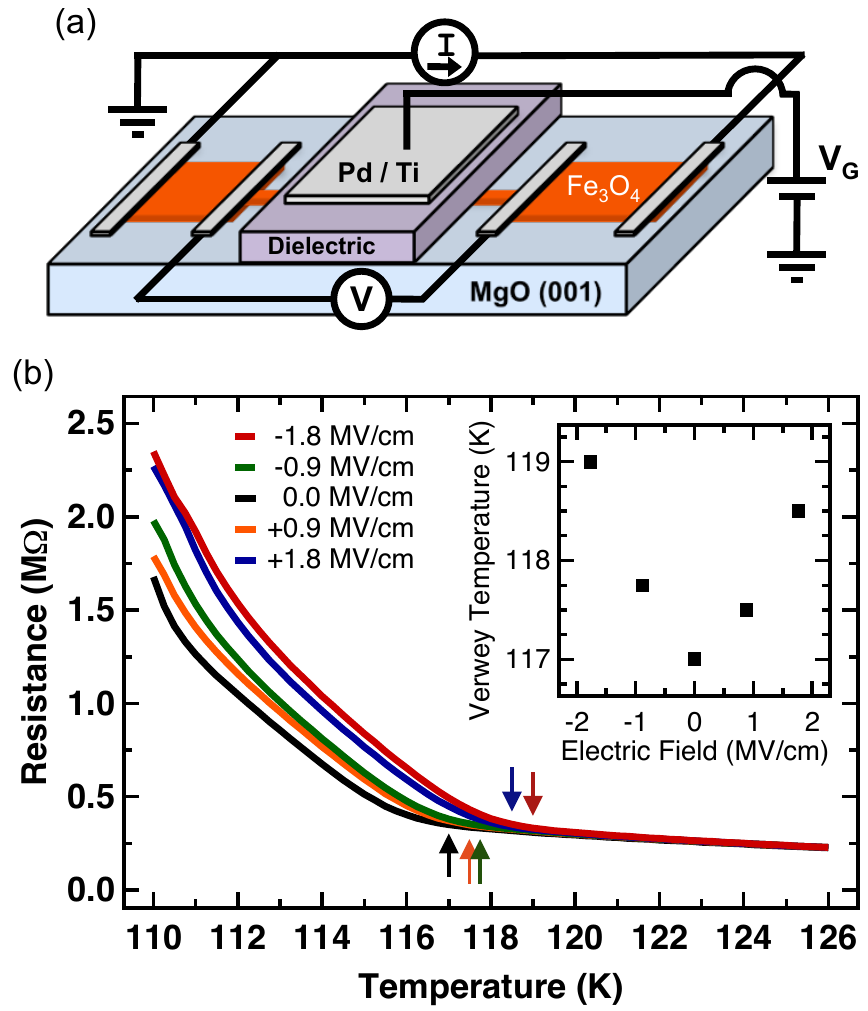}
\caption{\label{fig:tdep} Electrical gating of Fe$_3$O$_4$
and manipulation of the Verwey transition.  
(a) A schematic of the sample device structure.  The dielectric layer consists of 
PMMA(900 nm)/Al$_2$O$_3$(50 nm)/MgO(10 nm). Positive electric field 
corresponds to the application of positive 
voltage to the top gate electrode.  (b) Temperature dependence of
 resistance for applied electric fields of +1.8 MV/cm (blue), +0.9 MV/cm (orange), 
 0 MV/cm (black), -0.9 MV/cm (green), -1.8 MV/cm (red).  The arrows show 
 $T_V$ for each electric field, which is summarized in the inset. }
\end{figure}

To apply electric fields to the Fe$_3$O$_4$ film, an insulating layer (PMMA/Al$_2$O$_3$/MgO) is 
deposited on top of the Fe$_3$O$_4$, followed by a metallic electrostatic gate (Pd/Ti), as 
shown schematically in Figure 2a. The devices are fabricated through 
several steps of evaporation using shadow masks. For the Fe$_3$O$_4$ layer, a narrow 
channel is produced with a width of 210 $\mu$m, creating a small active area to reduce 
the occurrence of pinholes and gate leakage. The Fe$_3$O$_4$ channel length is 4.2 mm 
and the gate length is 3.3 mm. Alternate samples with Fe$_3$O$_4$ films covering a large 
area of the substrate produce similar results \cite{supplemental}. 
Pd(100 nm)/Ti(15 nm) contacts (for four-point 
probe) are e-beam evaporated through a shadow mask
in a separate system.  Then a 10 nm MgO layer is grown on the Fe$_3$O$_4$ followed 
by a 50 nm  Al$_2$O$_3$ layer. PMMA is then spin coated onto the sample 
at 3000 rpm and cured under a vacuum environment at 170$^\circ$C. The spin coating and 
baking sequence is repeated three times giving a final PMMA layer thickness of 900 nm. Finally, a 
shadow mask is used to grow the Pd(100 nm)/Ti(15 nm) top gate electrode. Typical gate leakage 
is 0.5 nA for electric fields of $\pm$1.8 MV/cm.

An electric field is produced by applying a voltage ($V_G$) between the gate electrode 
and the Fe$_3$O$_4$ film. Figure \ref{fig:tdep}b shows the 
temperature dependence of resistance for 
applied electric fields of +1.8 MV/cm ($V_G$ = +60 V, blue), +0.9 MV/cm ($V_G$ = +30 V, orange), 
0 MV/cm ($V_G$ = 0 V black), -0.9 MV/cm 
($V_G$ = -30 V, green), -1.8 MV/cm ($V_G$ = -60 V, red) 
with corresponding colored arrows indicating $T_V$. The data clearly show that $T_V$ varies as 
a function electric field, as summarized in the inset of Fig. \ref{fig:tdep}c.  
At zero electric field, $T_V$ is 
117 K. Strikingly, both positive and negative electric fields cause $T_V$ to increase, indicating 
that the shift in $T_V$ depends primarily on the magnitude of electric field as opposed to its sign. 
The maximum effect is observed for -1.8 MV/cm, where $T_V$ increases to 119 K, 
giving $\Delta T_V$ = +2 K. At a similar electric field, the largest $\Delta T_V$ we observe in 
our study is $\Delta T_V$ = +6 K for a large area sample \cite{supplemental}. The 
increase of $T_V$ cannot be due to Joule heating because a heating artifact would appear 
as a reduction of $T_V$. We also rule out effects of irreversible sample change by measuring 
the zero electric field temperature dependence of resistance before and after taking the 
data in Fig. \ref{fig:tdep}b and no irreversible changes were observed.  Finally, we observe 
that temperature dependent resistance above $T_V$ does not change with applied 
electric field, which shows that the metallic phase is insensitive to electric field.

\begin{figure}
\includegraphics[width=80mm]{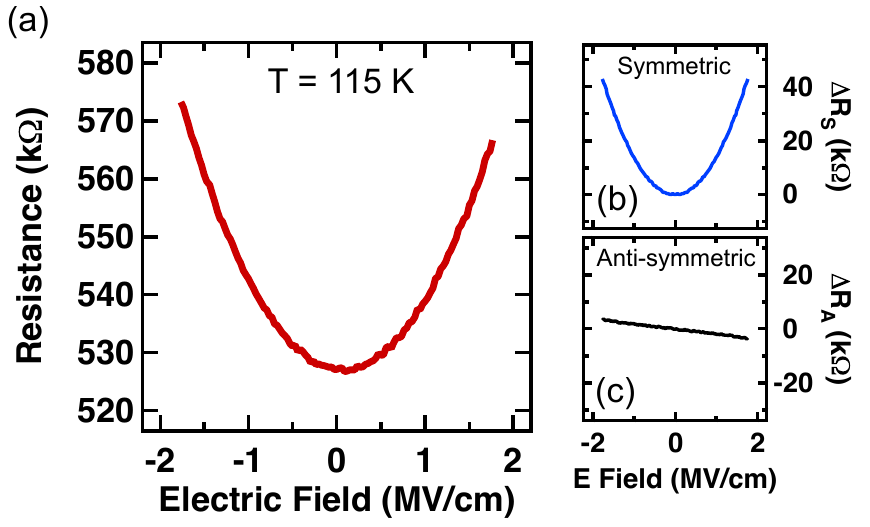}
\caption{\label{fig:gdep} Electrostatic gate dependence of resistance. 
(a) Gate dependent resistance for Fe$_3$O$_4$ at a temperature of 115 K.
(b) and (c) show the symmetric and anti-symmetric components of the gate 
dependent resistance change, respectively.  }
\end{figure}

To gain further insight into the electric field effect, we perform a detailed study of the 
gate dependent resistance under isothermal conditions. 
Figure \ref{fig:gdep}a shows the resistance at 
115 K as the electric field is swept between +1.8 MV/cm and -1.8 MV/cm.  Consistent with 
the shift in $T_V$ (inset Fig. \ref{fig:tdep}b), the resistance 
increases for both positive and negative electric 
fields and the effect is slightly larger for negative electric fields. To quantify the 
symmetry of the electric field effect, we separate 
the change in resistance $\Delta R(E) = R(E) - R(0) $
into a symmetric part $\Delta R_{S} (E) = [\Delta R(E) + \Delta R(-E)]/2$ (Fig. 3b) and anti-symmetric 
part $\Delta R_{A}(E) = [\Delta R(E) - \Delta R(-E)]/2$ (Fig. 3c). 
Comparing Fig. \ref{fig:gdep}b and \ref{fig:gdep}c, the symmetric 
part is up to 11 times larger than the anti-symmetric part. Because the change in 
carrier concentration is proportional to $E$ (i.e. anti-symmetric), the small contribution of $\Delta R_A$ indicates that electric field control of the Verwey transition is not driven by a carrier 
concentration effect. Instead, a symmetric effect can be driven by other interactions 
with the electric field. The presence of an electric field will induce electric polarization 
given by $P = \chi_{e} E = (\kappa-1)\epsilon_{0}E$, where $\chi_{e}$ is 
the electric susceptibility, $\epsilon_{0}$ is the permittivity 
of free space, and $\kappa$ is the relative dielectric 
constant of Fe$_3$O$_4$. The induced polarization 
will produce an energy contribution 
$U = -\frac{1}{2}(PE) =-\frac{1}{2}(\kappa -1)\epsilon_{0}E^2$ that is symmetric in $E$. In 
addition, as Fe$_3$O$_4$ undergoes the Verwey transition, the dielectric constant changes 
sharply with $\kappa$ being larger for the insulating state than for the metallic 
state $(\kappa_{ins} > \kappa_{metal})$ \cite{pimenov:2005}. Thus, energy is lower 
for the insulating state than 
for the metallic state, which stabilizes the low temperature insulating state and 
causes $T_V$ to increase. Therefore, this provides a macroscopic explanation for an 
electric field effect that is symmetric in $E$ and produces an increase in $T_V$, consistent 
with experimental results. Further theoretical work is needed, including a microscopic 
model that can provide an explanation for the magnitude of the effect. In addition, a 
contribution from electric field induced strain could generate this symmetry and 
should also be investigated \cite{nagasawa:2007}.

\begin{figure}
\includegraphics[width=60mm]{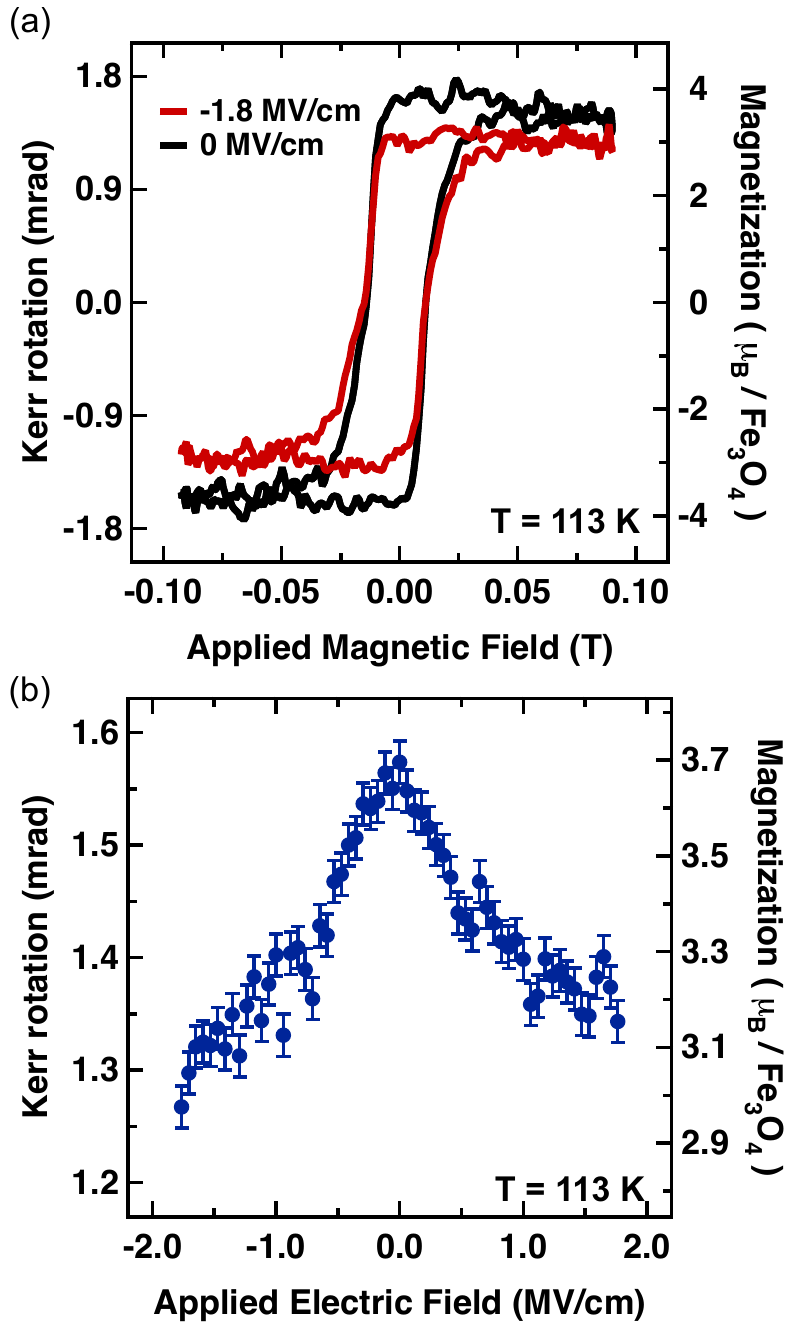}
\caption{\label{fig:me} Electrostatic gate dependence of magnetization. 
(a) MOKE loops measured at 113 K with applied electric fields of 0 MV/cm (black) 
and -1.8 MV/cm (red), showing a decrease in magnetization with the application 
of an electric field.  (b) Magnetization as a function of electric field, 
demonstrating a magnetoelectric effect induced by electric field control of the Verwey transition.}
\end{figure}

Since the Verwey transition in Fe$_3$O$_4$ is a correlated phase transition that couples both 
the charge and magnetic properties, it should be possible to tune magnetic properties 
with applied electric field. Figure \ref{fig:me}a shows MOKE 
hysteresis loops measured at 113 K with 
applied electric field of 0 MV/cm (black) and -1.8 MV/cm (red). The absolute 
magnetization is determined by SQUID measurements (right axis of Figure \ref{fig:me}a). 
An electric field of -1.8 MV/cm causes a decrease in the saturation magnetization of 18\%. 
Figure \ref{fig:me}b displays the saturation magnetization as the electric field is swept 
between +1.8 MV/cm and -1.8 MV/cm. With the application of either positive or negative 
field, the magnetization decreases with a slightly stronger effect for negative fields. 
The magnetoelectric behavior is generated because the magnetization has strong 
temperature dependence below $T_V$ (Figure \ref{fig:materials}d). 
When electric field is applied, the increase 
of $T_V$ causes magnetization M to decrease because $dM/dT$ is positive 
at T = 113 K; the intuitive picture is that the $M$ vs $T$ curve of Fig. \ref{fig:materials}d 
shifts in temperature 
as $T_V$ increases. Thus, the decrease of magnetization for both positive and negative 
fields (with slightly stronger effect for negative fields) is consistent with the electric 
field dependence of $T_V$ 
(Fig. \ref{fig:tdep}b inset) and resistance (Fig. \ref{fig:gdep}). The change in magnetization 
as a function of electric field is quantified by a magnetoelectric coefficient, 
$\alpha_{ME} =  | \Delta M/\Delta E |$, where $\Delta M$ is the 
change in magnetization and $\Delta E$ is the change in 
electric field.  Comparing the values at $E=0$ MV/cm and $E = -1.8$ MV/cm yields a 
value of $\alpha_{ME} = 585\pm39$ pT m/V.  This is about an order of magnitude larger than in 
other single-phase magnetoelectric materials, but not as large as some composite 
systems \cite{supplemental}. 
These results are compelling because they demonstrate a new method 
for generating magnetoelectric effects by controlling a correlated phase transition. 

In conclusion, we have demonstrated the electric field control of the Verwey transition 
in Fe$_3$O$_4$ thin films. An electric field stabilizes the charge-ordered insulating state 
causing the Verwey transition temperature to increase. By manipulating a correlated 
phase transition that combines both charge and magnetic transitions, we realize a large 
and novel magnetoelectric effect.

We acknowledge K. M. McCreary and P. M. Odenthal for their assistance. This material 
is based on research sponsored by DARPA/Defense Microelectronics Activity (DMEA) 
under agreement number H94003-10-2-1004. We acknowledge Youli Li at the UCSB MRL 
Central Facilities (NSF Award No. DMR 1121053) for technical assistance with x-ray 
diffraction measurements and discussion.

\bibliography{verwey_prl.bib}

\end{document}